\documentclass{aa}  

\usepackage{graphicx}
\usepackage{txfonts}

\usepackage{color, colortbl}
\usepackage[hidelinks]{hyperref}
\usepackage{xcolor}
\hypersetup{
    colorlinks = true, 
    linkcolor={red!50!black},
    citecolor={blue!50!black},
    urlcolor={blue!80!black}
}
\begin{document} 

\definecolor{bubbles}{rgb}{0.91, 1.0, 1.0}
\definecolor{columbiablue}{rgb}{0.61, 0.87, 1.0}
\definecolor{cream}{rgb}{1.0, 0.99, 0.82}
\definecolor{lightblue}{rgb}{0.68, 0.85, 0.9}
\definecolor{lightcyan}{rgb}{0.88, 1.0, 1.0}

   \title{Probing galactic double-mode RR Lyrae stars against Gaia EDR3}

   \author{Geza Kovacs
          \inst{1}
	  and 
	  Behrooz Karamiqucham
          \inst{1}
          }

   \institute{Konkoly Observatory, Budapest, 1121 Konkoly Thege ut. 15-17, Hungary\\
              \email{kovacs@konkoly.hu}
             }

   \date{Received April 15, 2021; accepted ??, 202?}


%
%
  \abstract
{Classical double-mode pulsators (RR~Lyrae stars and 
$\delta$~Cepheids) are important for their simultaneous pulsation 
in low-order radial modes. This enables us to put stringent 
constraints on their physical parameters.}
{We use $30$ bright galactic double-mode RR~Lyrae (RRd) stars 
to estimate their luminosities and compare these luminosities 
with those derived from the parallaxes of the recent data 
release (EDR3) of the Gaia survey.}
{We employ pulsation and evolutionary models, together with 
observationally determined effective temperatures to derive  
the basic stellar parameters.} 
{Excluding $6$ outlying stars (e.g., with blending issues) the 
RRd and Gaia luminosities correlate well. With the adopted 
temperature zero point from one of the works based on the 
infrared flux method, we find it necessary to increase the 
Gaia parallaxes by $0.02$~mas to bring the RRd and Gaia 
luminosities into agreement. This value is consonant with 
those derived from studies on binary stars in the context of 
Gaia. We examine also the resulting period-luminosity-metallicity 
(PLZ) relation in the 2MASS K band as follows from the RRd 
parameters. This leads to the verification of two independently 
derived other PLZs. No significant zero point differences are 
found. Furthermore, the predicted K absolute magnitudes agree 
within $\sigma=0.005$--$0.01$~mag.}
   {}


\titlerunning{Probing galactic double-mode RR Lyrae stars}
\authorrunning{Kovacs, G. \& Karamiqucham, B.}

   \maketitle
%
%
%
\section{Introduction}
\label{sect:intro}
The historical discovery of the first double-mode RR Lyrae star, 
AQ~Leo, in the galactic field by \cite{jerzy1977} and the 
subsequent discovery of 10 additional objects by \cite{cox1983} 
in the globular cluster M15, opened the possibility of deriving 
masses of RR~Lyrae stars directly from linear pulsation models, 
independently of stellar evolution theory. This is an important 
step in understanding the mass distribution of horizontal branch 
(HB) stars that results from the poorly known mass loss events 
in the thermonuclear instability phase during the final period 
on the first ascent to the giant branch, before falling down to 
the zero-age HB (ZAHB). 

Double-mode pulsation (i.e., simultaneous pulsation in low-order 
radial modes) was known among Cepheids well before 
1977 \citep{oosterhoff1957a,oosterhoff1957b}, and the first 
theoretical investigations quickly indicated a serious 
discrepancy between the pulsation and evolutionary masses 
\citep{petersen1973,stobie1977}. The solution of this discrepancy 
was not possible up until 1992, when stellar opacities were 
critically revisited by the opacity projects OPAL and OP 
\citep{iglesias1991,iglesias1996,seaton1994}. 
These studies nicely confirmed the `heretic' suggestion of 
\cite{simon1982}. He recognized that an increase of several 
factors in the heavy element opacities should solve the 
`beat Cepheid mass discrepancy' and also two other big issues 
of stellar pulsation (the excitation of $\beta$~Cephei stars 
and the mass discrepancy of the bump Cepheids\footnote{Assuming 
that the Hertzsprung progression \citep{hertz1926} is caused 
by the $2:1$ resonance between the fundamental and $2^{nd}$ 
overtone modes, as first suggested by \cite{simon1976} and 
confirmed by several subsequent studies, e.g., by \cite{buchler1990}.}). 

Problems of similar severity for the double-mode RR~Lyrae (RRd) 
stars did not seem to exist. Indeed, using the updated metallicities 
\citep{cox1991,kovacs1991,kovacs1992} did not alter the belief 
that these stars are far less sensitive to even the large opacity 
changes predicted by OPAL/OP. This finding, is, of course, not 
terribly surprising, due to the overall low metallicity of 
RR~Lyrae stars.  

In spite of these encouraging events, there are several fundamental 
questions still left unanswered concerning double-mode pulsations 
both in Cepheids and RR~Lyrae stars. Two issues stand out from 
these questions. The first one is the inability of the nonlinear 
hydrodynamical models to produce stationary double-mode pulsation 
for sound physical settings. Although there were reports in the 
literature of achieving this goal \citep{kovacs1993,kollath2002}, 
in our view \citep[see also][]{smolec2010} they were more of the 
results of fine tuning certain parameters and accidentally catch 
some models that showed some level of similarity to those that 
are actually observed. As a result, they do not provide clear 
pieces of evidence for the underlying source of sustained 
double-mode pulsation. This failure of nonlinear hydrodynamics 
extends also to the no-clue nature of the (quasi-)periodic 
modulation of the RR~Lyrae stars 
\citep[known as Blazhko effect,][]{blazhko1907}. 
The second issue is the increasing number of `strange' 
secondary periods both in Cepheids and RR~Lyrae stars 
\citep{moskalik2009,jurcsik2015}. These components, albeit 
with small amplitudes, are clearly identified in many cases 
both from space- and ground-based data. None of the radial 
eigenmode periods fit to the observations 
\citep{dziembowski2016,smolec2017}. 
Coupled with the other unsolved dynamical issues above, we have 
to admit that our knowledge on the intricate physical nature of 
these objects is just as limited as it was at the time when they 
were discovered.    

Against all odds, it seems as attractive now as in the early 
days to assume that the `classical' double-mode variables are, 
indeed, fundamental- (FU) and first overtone- (FO) mode pulsators 
and that the (highly precise) observed periods are close to 
the model periods derived from linear non-adiabatic (LNA) 
pulsation models. In this simplest form of stellar seismology, 
the standard approach utilizes the period -- period ratio diagram 
\citep[the so-called `Petersen' diagram,][]{petersen1973,petersen1978}  
to gain information on the stellar masses with the help of some 
knowledge on the heavy element metallicity, sometimes accessible 
by well-calibrated metallicity indicators 
\citep[e.g., Preston's $\Delta S$ index --][]{preston1959}. 

However, even assuming that the relative abundances of the 
species contributing to the overall heavy element abundance are  
fixed, the radial mode periods depend on four parameters: 
effective temperature, $T_{\rm eff}$, mass $M$, luminosity $L$ 
and metal abundance. Therefore, to aim for a more precise 
determination of $M$ and $L$ from the pulsation equations, in 
a series of earlier works we used also the observed color 
information to get a precise-enough proxy for $T_{\rm eff}$ 
\citep{kovacs1999,kovacs2000a,kovacs2000b}.    

Verification of the stellar properties derived from double-mode 
variables is also very important for the consistency of the method 
and for the validity of the basic assumptions on the observed 
periods. Earlier, this check was made through the comparison with 
cluster/galaxy distances \citep{kovacs1999,kovacs2000a,kovacs2000b} 
and period-luminosity-color relations based on Baade-Wesselink 
analyses \citep{dekany2008}. Now, with the Early Data Release 3 
\citep[EDR3, ][]{lindegren2021} of the Gaia mission, we have a 
new and exciting opportunity to verify the derived luminosities 
almost directly and independently of the stellar pulsation method. 
This approach was not feasible before EDR3, because of the factor 
of two or even larger errors on the DR2 parallaxes. 

In this paper we investigate the compatibility of the luminosities 
derived from the pulsation/evolution analysis of $30$ galactic 
field RRd stars with those computed from the EDR3 parallaxes. 
Furthermore, we also compare the period-luminosity-metallicity 
(PLZ) relation derived from the RRd pulsation with those obtained 
by other methods and using different datasets. 

%
%
\begin{table*}[t!]
\centering
\begin{minipage}{200mm}
\caption{The 30 galactic RRd stars analyzed in this paper}
\label{tab_obs_ref}
\scalebox{1.0}{
\begin{tabular}{lrlrlrlrlr}
\hline
Target    & Ref. & Target & Ref. & Target & Ref. & Target & Ref. & Target & Ref.\\
\hline 
V0500 Hya & 6& V5644 Sgr      & 6& V0374 Tel & 2& V0416 Pav & 3& J211848-3430.4 & 1\\
V0372 Ser & 4& NN Boo         & 7& AZ For    & 2& CU Com    &10& V0633 Cen      &12\\
Z Gru     & 3& BS Com         & 3& CZ Phe    & 2& CR Cap    & 2& QW Aqr         & 2\\
XX Crv    & 2& V0363 Dra      & 8& V0338 Boo & 8& AG PsA    & 2& J040054-4923.8 & 1\\
V0381 Tel & 6& SW Ret         & 1& V0458 Her & 8& V2493 Oph & 5& J141539+0010.1 & 1\\
AQ Leo    &13& AL Vol         & 6& XY Crv    & 9& BN UMa    &11& CF Del         &11\\
\hline
\end{tabular}}
\end{minipage}
\begin{flushleft}
\vspace{-5pt}
{\bf Notes:}
(1)~\cite{szczygiel2007} ; 
(2)~\cite{bernhard2006};
(3)~\cite{wils2006} (see also authors' note on an earlier mentioning 
BS~Com as an RRd star by Bragaglia  et al.)
(4)~\cite{garcia2001};
(5)~\cite{garcia1997} ;
(6)~\cite{wils2005} ;
(7)~\cite{koppelman2004} ;
(8)~\cite{wils_etal2006} ;
(9)~\cite{pilecki2007} ;
(10)~\cite{clementini2000} ;
(11)~\cite{clusky2008} ;
(12)~\cite{wils2010} ;
(13)~\cite{jerzy1977} 
\end{flushleft}
\end{table*}
%

%
%
\section{Datasets}
\label{sect:data}
We made a thorough search in the literature for relatively bright 
galactic field RRd stars with well-documented discovery analysis. 
We found $30$ objects viable for gathering further data to conduct 
pulsation and stellar evolution analysis and compare the derived 
luminosities with those obtained from the Gaia parallaxes. 
Table~\ref{tab_obs_ref} lists these stars, together with 
the references to the respective discovery papers.   

The stars are ordered in decreasing brightness, starting with 
$V=10.8$~mag for V0500~Hya and ending with $V=14.4$~mag for 
CF~Del. Most of the objects are between $V=12.5$ and $13.5$ 
magnitudes. The Gaia EDR3 parallaxes peak for the two brightest 
stars at $\sim 0.8$~mas and decline for the rest from $\sim 0.4$~mas 
to $0.15$~mas with a concomitant relative errors from $6$\% to 
$13$\%. The relative errors for the two brightest stars are in 
the ballpark of $3$\%, which is quite remarkable. All these show 
an impressive factor of two or even greater increase in the  
precision of the EDR3 parallaxes with respect of DR2.  

To employ RRd stars in estimating their luminosities, from the 
observational side, we need metallicity estimates and two-color 
photometry, free from interstellar reddening. Unfortunately, 
metallicities are not accessible for our targets. Therefore, we 
resort to the combination of the pulsation and stellar evolution 
models as described in Sect.~\ref{sect:method}. With respect 
to data accessibility and quality, we opt to choose the All Sky 
Automated Survey \citep[ASAS,][]{pojmanski1997} to estimate 
average magnitude in the Johnson V filter. For stars without 
ASAS photometry, we use Gaia photometry. To secure the 
$T_{\rm eff}$ estimate, we use the V photometry in combination 
with the infrared fluxes collected by the WISE satellite in 
the W1 and W2 bands. For the estimation of the reddening we 
employ the map of \cite{schlafly2011} accessible at the NASA 
IPAC site.\footnote{\url{https://irsa.ipac.caltech.edu/applications/DUST/}}
Additional details on the photometric datasets and the calibration 
methods employed are given in the following subsections.  

%
%
\subsection{The <V> magnitudes}
\label{sect:Vmag}
Thanks to the large sky coverage and long-term dedication of ASAS, 
$26$ stars from our target list have abundant time series. This 
allows us to perform stable Fourier fits and thereby to estimate 
accurately the magnitude average as the constant term $A_0$ 
in the Fourier representation
%
%
\begin{eqnarray}
\label{eq_four_sum}
X(t) &=& A_0 + \sum_{j=1}^{12} A_j \sin(2\pi\nu_j(t-t_0)+\phi_j) \hspace{2mm} , 
\end{eqnarray}
where \{$\nu_j$\} is the linear combination of the fundamental 
($f_0$) and first overtone ($f_1$) frequencies: 
$\nu_j=|n_0\times f_0+n_1\times f_1|$ with $n_0$ and $n_1$ satisfying 
the conditions $|n_0|\geq 0$,\, $|n_1|\geq 0$, and $0<|n_0|+|n_1|\leq 3$. 
In fitting Eq.~\ref{eq_four_sum} to the observations, first the 
time series \{$V$\} given in Johnson V magnitudes are converted to 
fluxes \{$F$\} via $F(t)=10^{-0.4V(t)}$. Once \{$F$\} is fitted 
with \{$X$\},the flux average $A_0$ is converted back to magnitudes 
simply by setting $<V>=-2.5\log A_0$. To get a better estimate 
on the average, outliers in the fit are omitted at the $3\sigma$-level. 
We note that the somewhat awkward procedure of transforming 
magnitudes back and forth is generally employed in computing the 
average light level of RR Lyrae stars, because the flux 
averages are closer to the static stellar model values 
than the magnitude averages \citep{bono1995}.   

%
%
\begin{figure}[h]
\centering
\includegraphics[width=0.40\textwidth]{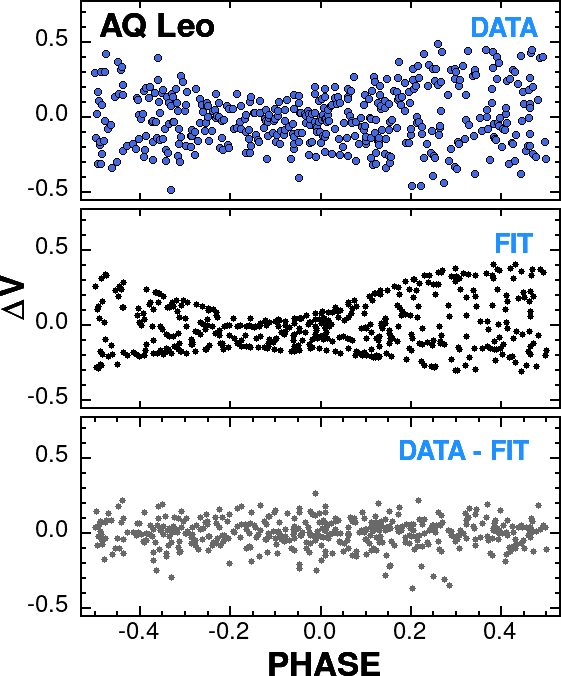}
\caption{{\em Upper panel:} ASAS light curve of AQ~Leo folded with 
         $1.615$~d, corresponding to the beat period of the 
	 FU and FO components of the pulsation. 
	 {\em Middle panel:} As above but for the $3^{\rm rd}$-order 
	 (12-component) Fourier fit given by Eq.~\ref{eq_four_sum}. 
	 {\em Lower panel:} Residuals of the Fourier fit.}
\label{AQ_Leo_fit}
\end{figure}

As an example of the type of time series we are dealing with, 
the light curve of the prototype of this class of RR~Lyrae 
stars, AQ~Leo is shown in Fig.~\ref{AQ_Leo_fit}. Because the 
two periods are incommensurate, the pulsation, strictly speaking, 
is non-periodic. Nevertheless, the period ratio is quite close 
to $3/4$, therefore, when folded with the beat period of $1.615$~d, 
we get a reasonably periodic pattern. The standard deviation of 
the $12$-component fit is $0.087$~mag, which is $\sim 50$\% 
larger than the formal errors. However, the formal errors are 
image errors, whereas the residual scatter is time series 
approximation errors, and the two quantities, albeit broadly 
correlated, are not the same. In any case, the over $400$ data 
points shrink the statistical error of the mean magnitude down 
to $0.0044$~mag, which is sufficiently accurate for our purpose.  

In this demonstration and throughout this work we chose the 
magnitudes obtained by the smallest aperture. The ASAS photometric 
pipeline uses $5$ aperture sizes with the smallest being only 
2 pixels wide (i.e., 30", due to the large field of view of 
$8.5^{\circ}\times 8.5^{\circ}$ on a $2K\times2K$ CCD 
chip.\footnote{\url{http://www.astrouw.edu.pl/asas/explanations.html}} 
Due to the low brightness of most of our RRd targets, blending is 
quite common, especially at larger apertures. Therefore, we choose 
the smallest aperture to minimize the effect of blending on the 
estimation of the average magnitude (even if other apertures yield 
light curves of lower scatter).  

%
%
\begin{figure}[h]
\centering
\includegraphics[width=0.45\textwidth]{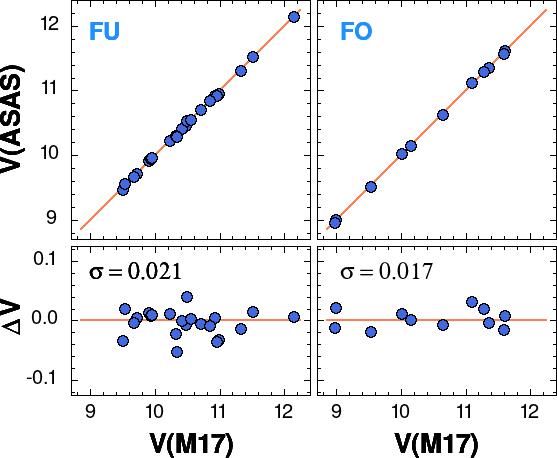}
\caption{Comparison of the ASAS average <V> magnitudes with those 
         of \cite{monson2017}. 
	 {\em Upper panels:} <V> vs <V> plots for the FU and FO 
	 variables.   
	 {\em Lower panels:} \cite{monson2017} <V> minus ASAS <V> 
	 as a function of <V> of \cite{monson2017}. Continuous 
	 lines indicate levels of equality.}
\label{asas_monson}
\end{figure}

It is important to ensure that the derived ASAS <V> magnitudes 
are compatible with high-precision observations. To test this, we 
rely on the recent compilation 
of \cite{monson2017}, including both archival data and new 
observations made by the Three-hundred MilliMeter Telescope (TMMT) 
of the Carnegie Observatories. By cross-matching their Table~5 and 
the ASAS archive, we find $22$ RRab and $11$ RRc stars common. 
The <V> magnitudes from these two sources are plotted in 
Fig.~\ref{asas_monson}. We see that the ASAS photometry does not 
show any sign of systematic offset for this relatively bright set 
of objects. The median differences and the mean errors are 
$+0.0017\pm 0.0045$ and $+0.0029\pm 0.0047$ for the FU and FO 
variables, respectively. 
  
As mentioned earlier, from the $30$ galactic RRd stars in our 
sample $26$ have ASAS data available. For the remaining $4$ 
stars we should find some way to estimate <V>. The obvious 
choice may seem to be some mixture of the Gaia magnitudes in 
the three bands. By using the FU/FO variables of \cite{monson2017}, 
we find that this does not introduce much of an improvement in 
the quality of the fit as compared with the single-color fit in 
the BP band only. Consequently, we use a simple linear transformation 
of the Gaia EDR3 BP magnitudes to estimate the corresponding 
flux-averaged V magnitudes. It is important to emphasize that 
the BP magnitudes used throughout this paper are 
{\em simple averages} (i.e., not the results of Fourier fits). 
Therefore, as we will see below, for large-amplitude, asymmetric 
light curves the transformation will likely yield erroneous 
estimates for <V>.   

There are $18$ FO variables with precise <V> magnitudes in 
the paper by \cite{monson2017}. We add the RRd stars AQ~Leo 
and BS~Com to this set, since both have non-ASAS-based average 
V magnitudes \citep{jerzy1982, dekany2008}. Because the published 
averages for the above two stars are magnitude averages, we 
subtract $0.015$ from the published values to get a rough estimate 
on the flux-averaged V magnitude \citep{bono1995}. We note 
that for a third RRd star, V0372~Ser, we have also independent, 
and assumed to be precise average V magnitude by \cite{benko2009}.  
However, their value seem to be discordant for some reason both with 
the ASAS value and with the Gaia-transformed value below. 
Consequently, we do not use this star in deriving our calibration 
formula. Finally, we use  $20$ <V> values in the calibration of 
the BP magnitudes. Employing robust least squares fit, we derive 
the following formula

%
%
\begin{eqnarray}
\label{eq_BP2V}
<V> = 10.972\pm 0.006 + (1.011\pm 0.006)(BP-11.097) \hspace{2mm}. 
\end{eqnarray}
Here we subtracted the average of BP over the sample, to have 
non-correlated errors on the regression coefficients. This 
transformation leads to a fit with $\sigma=0.022$~mag for the 
calibrating FO stars. 

%
%
\begin{figure}[h]
\centering
\includegraphics[width=0.45\textwidth]{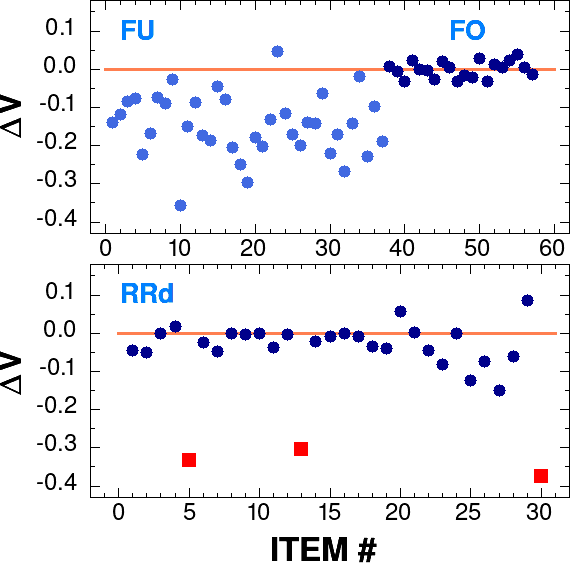}
\caption{{\em Upper panel:} <V> magnitudes of \cite{monson2017} 
         minus those calculated from the Gaia 
	 average BP magnitude (see Eq.~\ref{eq_BP2V}).
	 {\em Lower panel:} As above but for the ASAS <V> 
	 magnitudes of the RRd stars of this paper. Squares 
	 show the three outliers whose <V> magnitudes are 
	 seriously flawed (most likely by blending) in the 
	 ASAS setting -- see text for further details.}
\label{asas_gaia}
\end{figure}

Figure~\ref{asas_gaia} shows the striking difference 
between the FU and FO stars. As already mentioned above, this 
is primarily due to the fact that FU stars have highly 
non-sinusoidal light variation, and this makes the simple averages 
(BP average magnitudes) systematically fainter than the Fourier 
averages. For the RRd stars (lower panel of Fig.~\ref{asas_gaia}), 
a similar effect is observable, albeit at a much lower level. 
We note however, that less severe blending has the same effect, 
therefore, it is difficult to separate the two effects in the 
present case. Anyway, for the RRd stars, the overall shift remains 
in the range of a few hundredths of magnitude for the brighter part 
of our sample (the item ordering is also brightness ordering for 
the RRd sample). For fainter stars, the difference between the 
BP estimated <V> values and those obtained directly from the 
ASAS data becomes larger and noisier, most likely due 
to the more severe blend contamination at fainter magnitudes. 

We have three strong outliers among the RRd stars. The brightest 
star, V0381~Tel has, indeed a nearby bright companion, well within 
the $15"$ radius of the smallest aperture used by ASAS. In the 
case of V0374~Tel we have also a crowded field with two similarly 
bright companions some $20"-30"$ apart but it is unclear how do 
they affect the measured brightness of the target if we assume 
that the $15"$ aperture is correctly set. One possibility is that 
it is positioned on the photocenter of the poorly resolved stellar 
triplet (target and the nearby companions), and this leads to 
the higher flux for the target. The faintest star, CF~Del has 
only $34$ data points in the ASAS database, leading to an 
inaccurate estimate of the average magnitude (the field for this 
target is also crowded but the brighter companions are out of 
the smallest aperture we use).  

In summary, according to the analysis presented in this section, 
we use the Fourier-based flux-averaged magnitudes of the ASAS 
observations for $23$ stars. For the remaining $4+3=7$ stars we 
estimate <V> from the BP magnitudes of the EDR3 of the Gaia mission. 
We refer to Table~\ref{tab_obs_dat} for the actual values used. 

%
%
%
\subsection{The K magnitudes}
\label{sect:Kmag}
The method employed in this paper to derive the parameters of 
double-mode star, requires the knowledge of the effective 
temperature. Thanks to the all-sky survey of the WISE infrared 
satellite \citep{wright2010}, we have accurate fluxes 
available for all of our targets. Together with the <V> 
magnitudes (as detailed in Sect.~\ref{sect:Vmag}), we can 
use one of the numerous $V-K\rightarrow T_{\rm eff}$ calibrations 
available in the literature. In order to do that, we need to 
convert the WISE W1 and W2 magnitudes to K magnitudes (Johnson 
or 2MASS, since they differ only by an additive constant: 
$K_{\rm Johnson} = K_{\rm 2MASS} + 0.03$)\footnote{See 
\url{https://old.ipac.caltech.edu/2mass/releases/second/doc/sec6_3.h} , 
counterchecked for this paper by the inspection of the data used in 
\cite{liu1989,liu1990}}. We find that the very recent 
$K_{\rm 2MASS}$ {\em flux-averaged} magnitudes of nearly $100$ 
galactic field RR~Lyrae stars by \cite{layden2019} and also 
the very recent catalog of the W1, W2 band-merged fluxes by 
\cite{schlafly2019} (the unWISE catalog) suit to this task 
very well.   

After some testing, we find that a simple linear formula 
does the job, and further complexities do not lead to any 
significant improvement. We find that the formula below 
fits the dereddened 2MASS magnitudes $K_0$ of \cite{layden2019} 
with a standard deviation of $0.033$~mag 

%
%
\begin{eqnarray}
\label{eq_W2K}
K_0 = 9.906\pm 0.006 + (1.003\pm 0.006)(W_0 - 9.906) \hspace{2mm}.  
\end{eqnarray}
Where $W_0$ denotes a mixture of the dereddened unWISE magnitudes 
computed from the published band-merged fluxes and reddening 
coefficients of \cite{wang2019} as follows

%
%
\begin{eqnarray}
\label{eq_W2K_plus}
W1   & = & 22.5 - 2.5\log FLUX(W1)     \hspace{2mm}, \nonumber \\
W2   & = & 22.5 - 2.5\log FLUX(W2)     \hspace{2mm}, \nonumber \\
(W1)_0 & = & W1   - 0.121E(B-V)        \hspace{2mm}, \nonumber \\
(W2)_0 & = & W2   - 0.081E(B-V)        \hspace{2mm}, \nonumber \\
W_0  & = & 0.3(W1)_0 + 0.7(W2)_0       \hspace{2mm}, \nonumber \\
K_0  & = & K_{\rm 2MASS} - 0.242E(B-V) \hspace{2mm}.
\end{eqnarray}

%
%
\begin{figure}[h]
\centering
\includegraphics[width=0.40\textwidth]{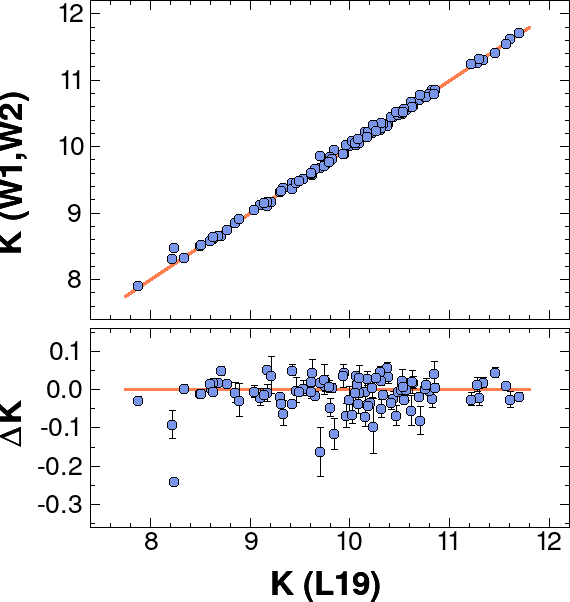}
\caption{Calibration of the unWISE \citep{schlafly2019} W1, W2 
         magnitudes to the 2MASS K magnitudes of the RR~Lyrae 
	 sample of \cite{layden2019}. On the vertical axis in the 
	 upper panel, dereddened magnitudes $K(W1,W2)=K0$ (see 
	 Eqs.~\ref{eq_W2K_plus}) are plotted. The reference 
	 levels are shown by continuous lines, $\Delta K=K(L19)-K(W1,W2)$. 
	 The strongest outlier in the lower panel is AR~Per, with 
	 an outstandingly large reddening of $E(B-V)=1.110$.}
\label{fig_calib_K}
\end{figure}

The quality of the calibration of the $98$ stars is exhibited in 
Fig.~\ref{fig_calib_K}. The error bars are solely from the errors 
of the K data, since the errors of $\sim 0.5$\% from the unWISE 
catalog are negligible. We see that, except for AR~Per, 
all, more substantially deviating stars have large error bars. 
AR~Per outlier status might be related to its outstandingly high 
reddening of $E(B-V)=1.110$. By using the extinction ratios of 
\cite{yuan2013} -- that are some $30$\%--$50$\% larger than those 
of \cite{wang2019} -- only exacerbates the outlier status of this 
star. One possibility is that there is a high inhomogeneity in 
the interstellar matter in the direction of AR~Per that happens 
to become much more transparent in an area that is not resolvable 
by the instrument which the map of \cite{schlafly2011} is based on. 
Another (perhaps more likely) possibility is that the foreground 
extinction is considerably lower than the total extinction given 
by the reddening map. 

%
%
\subsection{The $T_{\rm eff}(V-K)$ zero point}
\label{sect:teff_zp}
In our earlier studies \citep[e.g.,][hereafter KW99]{kovacs1999} 
we used the stellar atmosphere models of 
\citep[][hereafter C97]{castelli1997} 
to relate metallicity, gravity and color to the 
effective temperature, with the zero point tied to the 
`generally accepted' value based on the Infrared Flux Method 
(IRFM). Here we follow the same method but update the zero 
point by the one given for giants by \cite{gonzalez2009} 
(hereafter GB09). 

In tying the zero point, we encounter two problems. First, the 
calibration of GB09 for giants is limited to $T_{\rm eff} < 6300$~K, 
with only four points above $6000$~K. Therefore, the preferred 
$T_{\rm eff}$ range of $\sim 6500-7000$~K of the RRd stars is 
basically extrapolated if one relies on the formula of GB09. 
Second, the formula of KW99 is linear, whereas that of GB09 is 
nonlinear, exposing an issue of matching the two calibrations. 

%
%
\begin{figure}[h]
\centering
\includegraphics[width=0.40\textwidth]{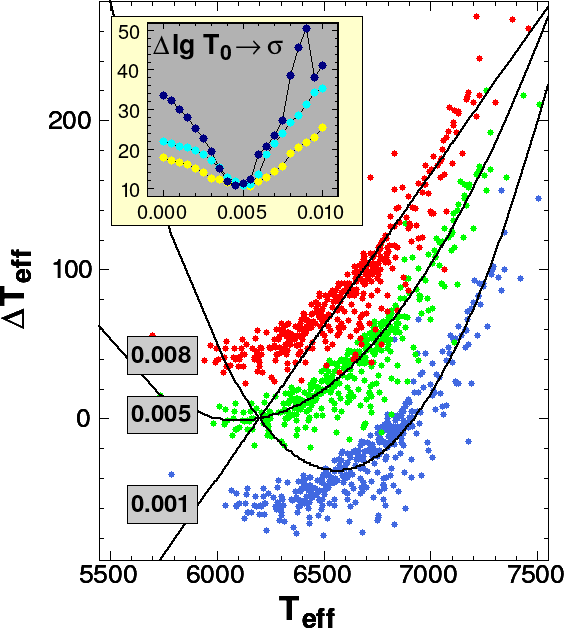}
\caption{Calibration of the ZP shift $\Delta\log T_0$ of the 
         $T_{\rm eff}$ formula (Eq.~\ref{eq_teff}). The vertical 
	 axis gives 
	 $\Delta T_{\rm eff}=T_{\rm eff}(GB09)-T_{\rm eff}$. 
	 The various ridges correspond to the ZP shifts given 
	 in the gray boxes. Continuous lines show the correction 
	 polynomial (Eq.~\ref{eq_teff_pol}) at an anchor temperature 
	 of $T_{\rm anc}=6200$~K. The inset displays the RMS 
	 of the polynomial-corrected residuals as a function 
	 of the ZP shift and $T_{\rm anc}$ (from bottom 
	 to up, for $T_{\rm anc}=6000, 6200, 6400$~K).} 
\label{teff_ZP}
\end{figure}

The linear formula presented by KW99 matches the C97 models with 
a residual standard deviation (in $\log T_{\rm eff}$) better than 
$0.001$ (corresponding to $15-20$~K). 
The parameter range covered by the fit is as follows: 
$6000 < T_{\rm eff} < 8000$~K, $2.5 < \log g < 3.5$, 
$-2.5 < [M/H] < -0.5$. Therefore, we rely on this high-quality 
fit concerning the parameter dependence. For matching the zero 
point (ZP) with that of BG09, we proceed as follows. 

First we choose a set of RR~Lyrae stars with reliable $V-K$ colors  
and metallicities. After testing various selections from the 
literature, we opt the abundant sample of \cite{dambis2013}, 
containing $383$ entries with the required number of parameters. 
To compare the $T_{\rm eff}$ estimates by the KW99 and GB09 formulae, 
for the KW99 formula we also need to know the gravity. For RR~Lyrae 
stars the static gravity can be evaluated fairly well even with a 
limited knowledge on the stellar mass. Using a linear approximation 
of the type of \cite{van_albada1971}, KW99 derived the following 
equation from their LNA models   

%
%
\begin{eqnarray}
\label{eq_logg}
\log g & = & 2.9383 + 0.2297\log M - 0.1098\log T_{\rm eff} \nonumber \\
       & - & 1.2185\log P_0 \hspace{2mm}.    
\end{eqnarray}
Without introducing any significant error (at least in the present 
context), one can fix the stellar mass $M$ at some `overall' value. 
We use $0.75M_{\odot}$ in this particular test. 

After computing the two $T_{\rm eff}$ estimates at any given ZP, 
we get a nonlinear dependence of 
$\Delta T_{\rm eff} = T_{\rm eff}(\rm GB09)-T_{\rm eff}$ 
on $T_{\rm eff}$ due to the nonlinearity of the GB09 formula 
(see Fig.~\ref{teff_ZP}). Here $T_{\rm eff}$ denotes  
the KW99 fit to the atmosphere models of C97 with the ZP shift 
of $\Delta\log T_0$ (see Eq.~\ref{eq_teff}). To compute the 
RMS at any given ZP, we need to eliminate the systematic 
difference, entirely due to the extrapolation of the GB09 
formula from the more populated, low-$T_{\rm eff}$ regime. 
The rectification is made by fitting a correction polynomial 
of order two to $\Delta T_{\rm eff}$

%
%
\begin{eqnarray}
\label{eq_teff_pol}
\Delta T_{\rm eff} = c_1 X + c_2 X^2 \ ,\ X = T_{\rm eff} - T_{\rm anc} \hspace{2mm}.    
\end{eqnarray}
Note that we do not have adjustable constant in the above equation, 
since this is given as a pre-selected ZP of the KW99 formula 
(see below). As a result, at the anchor temperature $T_{\rm anc}$ 
no polynomial correction is made to the residual. For testing 
the dependence of ZP on $T_{\rm anc}$, we select three anchors 
at the tail of the GB09 calibrating sample. The resulting ZPs 
should be quite independent of the actual values of these anchors. 

Our $T_{\rm eff}$ formula with the variable ZP shift of 
$\Delta\log T_0$ reads as follows

%
%
\begin{eqnarray}
\label{eq_teff}
\log T_{\rm eff}&=& 3.9158 - \Delta\log T_0 - 0.1156(V-K) + 0.0069\log g \nonumber \\
                &-& 0.0026[M/H] \hspace{2mm}, 
\end{eqnarray}
where $V$ and $K$ are in the Johnson system and reddening-free. 
The metallicity [M/H] is assumed to be solar-scaled, so that 
$[M/H]=[Fe/H]=\log Z/Z^{*}_{\odot}$, with 
$Z^{*}_{\odot}=0.02$.\footnote{The C97 models refer to the then 
available/accepted solar heavy element abundance. This is the 
reason for marking $Z_{\odot}$ by an asterisk.} 
We remind that $K_{\rm Johnson} = K_{\rm 2MASS} + 0.03$. 

For the anchor temperature of $6200$ we illustrate the pattern 
of the temperature difference $\Delta T_{\rm eff}$ in 
Fig.~\ref{teff_ZP}. The dots of various colors correspond to 
the ZP shifts $\Delta\log T_0$ as indicated in the gray boxes. 
All correction polynomials are zero at $T_{\rm anc}=6200$~K. 
For incorrect ZPs these polynomials provide bad fits 
to $\Delta T_{\rm eff}$. By scanning the possible ZPs, we end 
up with the plot shown in the inset. Independently of the value 
of $T_{\rm anc}$, we conclude with the same ZP shift of 
$0.0040$ -- $0.0055$. Finally, we settle at 
$\Delta\log T_0=0.0045$ in Eq.~\ref{eq_teff}.

%
%
%
\subsection{The bolometric correction}
\label{sect:bc_zp}
Because the theoretical models work with the total irradiated 
flux, we need to convert the luminosity to the wave-band-limited 
colors to utilize the observed magnitudes in constraining the 
models. This is done with the intermediation of the bolometric 
correction BC, and can be evaluated in various ways. As in our 
earlier works, we opt to the stellar atmosphere models of C97 
also in this study. For completeness, here we repeat the formula 
given in KW99 for BC with respect of the Johnson V filter  

%
%
\begin{eqnarray}
\label{eq_bc}
BC & = & 0.1924 + 0.0633u - 0.0411u^2 - 0.0233u^3 \nonumber \\
   & - & 0.0464\log g + 0.0689f + 0.0118f^2 - 0.0121fu \nonumber \\
f  & = & [M/H] \ , \ u = c(\log T_{\rm eff} - t) \nonumber \\
c  & = & 2/(\log T_2 - \log T_1) \ , \ t = (\log T_1 + \log T_2)/2 \nonumber \\
T_1& = & 6000 \ , \ T_2 = 8000 \hspace{2mm} .
\end{eqnarray}
This formula fits the model data with an RMS of $0.001$~mag in the 
parameter range of interest (see Sect.~\ref{sect:teff_zp}). It is 
recalled that the zero point of this formula has been shifted by 
$0.113$~mag with respect of the model values given by the C97 
models. With this shift, the models yield BC$_{\odot} = -0.082$, 
which is very close to $-0.07$, following from the current 
bolometric and Johnson V magnitudes of the Sun by \cite{willmer2018}. 
The bolometric magnitude $M_{\rm bol}$ of the Sun is fixed to 
$4.74$ throughout this paper, also in agreement with \cite{willmer2018}.   

To determine the luminosity, we proceed with the well-known basic 
formulae 
  
%
%
\begin{eqnarray}
\label{eq_lum}
M_{\rm V}         & = & <V> - \ R_{\rm V}E(B-V) + 5 + 5\log(0.001\pi) \nonumber \\
\log(L/L_{\odot}) & = & -0.4(M_{\rm V} - M_{\rm bol}{\odot} + BC) \hspace{2mm}, 
\end{eqnarray}
where the symbols have their standard meanings. The parallax $\pi$ is given 
in the units of milliarcsecond [mas].

%
%

%
%
\section{Brief description of the method}
\label{sect:method}
The basic idea of using double-mode stars in estimating their 
physical properties is the same as in our earlier works (e.g., KW99). 
However, due to the lack of abundance measurements for the targets, 
we need to involve also stellar evolution models, much in the same 
way as we did in \cite{dekany2008}. 

%
%
\begin{table}[h!]
\centering
\begin{minipage}{100mm}
\caption{Parameter ranges of the pulsation models}
\label{tab_puls_seq}
\scalebox{1.0}{
\begin{tabular}{lll}
\hline
Parameter     & Range/Values              & Step \\
\hline 
Mass          & $0.40$ -- $1.00$          & 0.05 \\
Luminosity    & $30$ -- $100$             & 5    \\
$T_{\rm eff}$ & $6000$ -- $8000$          & 100  \\           
Z             & 0.00001, 0.00002, 0.00003, 0.00005 &  -- \\ 
              & 0.00006, 0.00010, 0.00020, 0.00030 &  -- \\
              & 0.00044, 0.00060, 0.00079, 0.00100 &  -- \\
              & 0.00140, 0.00200, 0.00300, 0.01000 &  -- \\ 
\hline
\end{tabular}}
\end{minipage}
\begin{flushleft}
\vspace{-5pt}
{\bf Notes:} All models have Hydrogen abundance of $0.76$. 
\end{flushleft}
\end{table}
Concerning the pulsation models, we ran a new set of LNA models 
with a parameter range that is far wider than what is supposed 
to be occupied by the RRd stars. The model grid covers the full 
instability strip and beyond. The actual LNA model grid 
parameters are described in Table~\ref{tab_puls_seq}. All 
models have $400$ mass zones down to $5\times 10^6$~K. Although 
this number of zones could already provide an accurate 
estimates both for the period and the growth rates, nearly 
independently of the zoning, we followed the `historical' 
route of arranging the mass shells, largely inherited from 
\cite{stellingwerf1975} \citep[see also][]{kovacs1988}. This 
implies putting $80$ zones of equal mass from the surface, 
down to the Hydrogen ionization zone at $11000$~K. The remaining 
zones have geometrically increasing masses down to our core 
boundary at $5\times 10^6$~K. The code we use is identical 
with the one used in our earlier works, i.e., all of our models 
have pure radiative envelopes\footnote{Without the capability 
of modeling sustained double-mode pulsation, the relevance of 
nonlinearity and convection is unknown in the context of 
using periods for stellar parameter determination. Employing 
simple LNA periods does not seem to contradict any theoretical 
or observational constraint at this point.}
with the diffusion approximation, using the OPAL opacities 
\citep{iglesias1996}. 

\begin{figure}[h]
\centering
\includegraphics[width=0.40\textwidth]{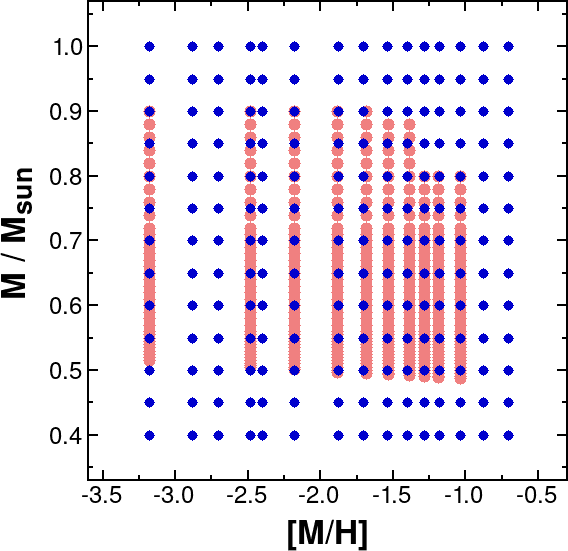}
\caption{Model metallicities vs mass for the grid values 
         of the BaSTI stellar evolution models (light 
	 coral) and the pulsation models (blue) employed 
	 in this paper. The metallicities are scaled with 
	 $Z_{\odot}=0.0152$ \citep{caffau2011}.} 
\label{fig_MZ_grid}
\end{figure}

For the stellar evolution input, we choose the recently updated 
models known as 
BaSTI\footnote{Bag of Stellar Tracks and Isochrones, 
\url{http://basti-iac.oa-abruzzo.inaf.it/index.html}} 
by \cite{hidalgo2018}. The models we use are Horizontal Branch 
(HB) models allowing evolution off the Zero Age HB up to the 
asymptotic giant branch. All models employed in this work are 
without $\alpha$ element enhancement\footnote{Some discussion 
of the effect of $\alpha$ enhancement can be found in 
Sect.~\ref{sect:LL}.}, convective overshooting and diffusion. 
For a comparison of the grid distributions in the metallicity 
-- mass space we show these grids for the pulsation and evolutionary 
models in Fig.~\ref{fig_MZ_grid}. Due to the large topological 
sensitivity of the evolution tracks at low- to medium masses, 
the evolutionary models are more densely sampled for the mass 
in this parameter regime. For the pulsation models we do not 
have such an effect, therefore, they are uniformly sampled for 
the mass, whereas taking values near and between the pre-given 
metallicities by the evolution models.  
 
%
%

\begin{figure}[h]
\centering
\includegraphics[width=0.40\textwidth]{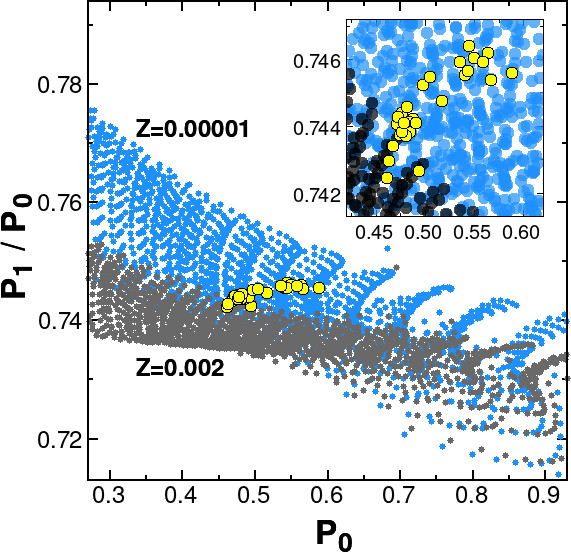}
\caption{Period -- period ratio diagram for the $30$ galactic 
         RRd stars investigated in this paper (yellow filled 
	 circles). The blue and gray dots show two sets of 
	 pulsation models for the metallicity labels 
	 (corresponding to $[M/H]=-3.18$ and $-0.88$ if 
	 $Z_{\odot}=0.0152$). In the main plot every second 
	 models are plotted, whereas in the inset all models 
	 fitting in the constrained period regime are shown.} 
\label{fig_petersen}
\end{figure}

Before proceeding with the description of the method of stellar 
parameter determination, in Fig.~\ref{fig_petersen} we quickly 
illustrate the position of the $30$ galactic RRd stars in the more 
customary $P_0$--$P_1/P_0$ plane \citep{petersen1973,petersen1978}.   
We chose pulsation models sandwiching the metallicity 
range of the RRd stars to indicate the minimum and maximum 
metallicities allowed by the observed periods. Without posing  
further constraints on the pulsation models, the Petersen 
diagram itself gives only a very rough estimate on the 
stellar parameters (primarily due to the degeneracy between 
the metallicity and mass). Still, the arc of the observed RRd 
stars in the $P_0$--$P_1/P_0$ plane clearly indicates the 
chemical inhomogeneity of the sample. We recall that this 
pattern was first recognized by the MACHO Collaboration in 
their work on the multimode RR~Lyrae inventory of the Large 
Magellanic Cloud \citep{alcock1997}.    

%
%

By using additional -- observational and theoretical -- pieces 
of information, we can place further constraints on the LNA 
models and arrive to a solution for the basic stellar parameters.   
In brief, we follow the steps below to derive the mass, luminosity 
and metallicity for the individual targets. 

\noindent
{\em Set input parameters:} 
Namely, periods of the FU and FO modes, 
$P_0$ and $P_1$; flux-averaged Johnson V magnitude (or, if it 
is not available, then Gaia BP magnitude [to be converted to 
Johnson V, via Eq.~\ref{eq_BP2V}]); unWISE W1 and W2 fluxes 
converted to 2MASS K magnitude via Eqs.~\ref{eq_W2K}, 
\ref{eq_W2K_plus}; reddening E(B-V). 

\noindent
{\em Select pulsation models:}
Interpolate all available LNA models to the $T_{\rm eff}$ value, 
given by the input parameters and generate dense $(M,L)$ grid 
from the models by quadratic interpolation on the logarithmic 
values. Select models satisfying the following period match 
constraint 
%
%
%
\begin{eqnarray}
\label{eq_dp}
\sqrt{\log^2 \Biggl({P_0^{\rm obs} \over P_0^{\rm lna}}\Biggr) + \log^2 \Biggl({P_1^{\rm obs} \over P_1^{\rm lna}}\Biggr)} \ < \ DP_{\rm max} \hspace{2mm} ,
\end{eqnarray}
where the upper limit of the logarithmic period distance 
$DP_{\rm max}$ is set equal to $0.1$\%. No interpolation is made 
to generate a dense metallicity grid. Finally, we get a dense 
$(M,L)$ grid, exactly matching the input temperature at various 
metallicities and periods satisfying Eq.~\ref{eq_dp}. 

\noindent
{\em Select evolutionary models:}
Interpolate the stellar evolution tracks linearly to the same 
temperature as for the pulsation models at any fixed metallicity 
$Z$. Then, generate a dense grid by interpolating the $(M,L)$ 
values corresponding to this fixed temperature. As for the 
pulsation models, no interpolation is made for the metallicities. 
We end up with densely sampled model grid for $(M,L)$, matching 
the observed temperature at different metallicities. 

\noindent
{\em Matching pulsational and evolutionary models:}
For the two $(M,L)$ sets (LNA and HB evolutionary models) we 
can search for the closest $(M,L)$ pairs at the same metallicity 
and accept as a solution by minimizing the simple distance 
measure given by 
%
%
%
\begin{eqnarray}
\label{eq_ml_match}
{\cal D}(M,L) = \sqrt{\log^2 \Biggl({L^{\rm ev} \over L^{\rm lna}}\Biggr) + \log^2 \Biggl({M^{\rm ev} \over M^{\rm lna}}\Biggr)} \hspace{2mm} .
\end{eqnarray}

\noindent
{\em Iteration on [Fe/H]:}
Because the metallicity plays a role in the estimation of the 
input parameters (i.e., $T_{\rm eff}$, and therefore $\log g$) 
We need to iterate on the solution to bring the starting 
metallicity and the one derived via Eq.~\ref{eq_ml_match} into  
close agreement. We find that within $\sim 10$ iterations one 
can reach the state of consistency between the input and output 
metallicities.  

\begin{figure}[h]
\centering
\includegraphics[width=0.40\textwidth]{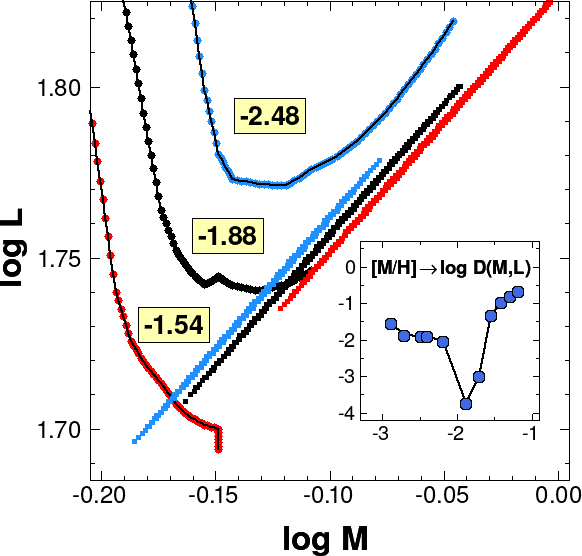}
\caption{Derivation of the mass, luminosity and metallicity 
         of AQ~Leo from the pulsational and evolutionary 
	 models. The labels show the corresponding metallicities 
	 with color coding applied also to the pulsation 
	 models. A shift of $\pm 0.003$ in $\log L$ is used 
	 for the pulsation models to ease the visibility of 
	 the metallicity effect (no shift for [M/H]$=-1.88$, 
	 $+0.003$ and $-0.003$ for [M/H]$=-2.48$ and $-1.54$, 
	 respectively). The inset shows the variation of the 
	 $(M,L)$ distance metric (Eq.~\ref{eq_ml_match}) as 
	 a function of the metallicity. All models have the 
	 same effective temperature of $6595$~K, and the 
	 metallicities are scaled with $Z_{\odot}=0.0152$.} 
\label{fig_ML_match}
\end{figure}

To illustrate the method at work, in Fig.~\ref{fig_ML_match} 
we show the position of the evolutionary and pulsation models 
at three different metal abundances for AQ~Leo, the prototype 
of the double-mode RR~Lyrae stars. All models have the same 
$T_{\rm eff}$ of $6595$~K as derived from the ASAS V and unWISE 
W1, W2 magnitudes. It is seen that the evolutionary and 
pulsational models exhibit opposite dependence on the metallicity, 
enabling a relatively secure selection of the metallicity 
range where the iso-$T_{\rm eff}$ curves cross.   

%
%
%

%
%
\section{Comparison with the Gaia EDR3 luminosities}
\label{sect:LL}
The early version of the Gaia DR3 enables us to perform a stringent 
test on the luminosities we derived from the RRd stars. The Gaia 
luminosities are basically independent of these luminosities, 
since only the bolometric correction contains a dependence on 
$T_{\rm eff}$, $\log g$ and [Fe/H], entering also in the evaluation 
of the RRd luminosities. Errors and RRd solution dependencies play 
a secondary effect in the actual value of BC. 

%
%
\begin{figure}[h]
\centering
\includegraphics[width=0.40\textwidth]{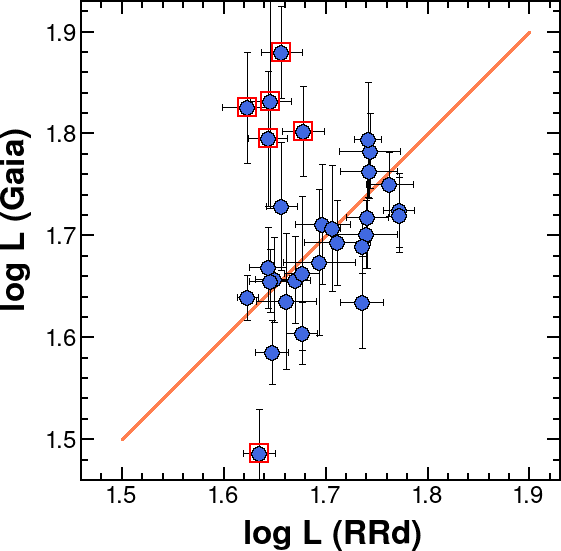}
\caption{Comparison of the luminosities derived for $30$ 
         galactic RRd stars (Tables\ref{tab_obs_ref}, 
	 \ref{tab_obs_dat}) and those computed from the 
	 Gaia EDR3 parallaxes. Outliers are marked by 
	 squares and discussed in the text. For reference, 
	 the equal luminosity values are shown by the 
	 orange line.} 
\label{fig_LL}
\end{figure}

The RRd luminosities are computed for all the $30$ targets as 
described in Sect.~\ref{sect:method}. For the evaluation of the 
EDR3 luminosities, we add $0.02$~mas to the EDR3 parallaxes (see 
later in this section for some reasonings of this correction). The 
luminosities are compared in Fig.~\ref{fig_LL}. All error sources 
(from BC, <V> and parallax) are included in the vertical error 
bars shown as $1\sigma$ limits for the Gaia luminosities. 
For the RRd luminosities the main error source is the zero point 
of the temperature scale. There are no other observational 
errors, since the periods are very accurate. However, there might 
be various/unknown errors in the theories we use. Since these 
errors are difficult to assess precisely enough, and both the 
evolutionary and the pulsation models proved to be quite accurate 
in various other applications, we pack all the possible errors 
in the known-to-be somewhat ambiguous zero point of $T_{\rm eff}$. 
We compute RRd luminosities with $\log T_{\rm eff}$ ZPs shifted 
by $\pm 0.005$ in respect of the ZP given in 
Sect.~\ref{sect:teff_zp}. Then, the horizontal error bars in the 
plot are given as half of the difference between the luminosities 
obtained with the two extreme $\log T_{\rm eff}$ ZPs.    

%
%
\begin{table*}[t!]
\centering
\begin{minipage}{200mm}
\caption{Derived parameters of 30 galactic RRd stars}
\label{tab_rrd_par}
\scalebox{1.0}{
\begin{tabular}{lccccccccccc}
\hline
 Target   & $\log L_G$ & $\log T_{\rm eff}$ & $\log M$ & $\log L$   & [M/H]            & Age        & DML \\
\hline\hline
V0500 Hya             & $ 1.764 \pm  0.027$ &  $3.8175$ & $-0.127 \pm  0.008$ & $ 1.742 \pm  0.028$ & $-1.88 \pm  0.50$ & $  61 \pm   16$ & $0.0003$ \\
V0372 Ser             & $ 1.639 \pm  0.022$ &  $3.8189$ & $-0.160 \pm  0.012$ & $ 1.622 \pm  0.011$ & $-1.18 \pm  0.05$ & $  25 \pm   14$ & $0.0379$ \\
Z     Gru             & $ 1.604 \pm  0.030$ &  $3.8289$ & $-0.170 \pm  0.008$ & $ 1.677 \pm  0.015$ & $-1.40 \pm  0.06$ & $  56 \pm   24$ & $0.0002$ \\
XX    Crv             & $ 1.751 \pm  0.031$ &  $3.8271$ & $-0.131 \pm  0.001$ & $ 1.762 \pm  0.023$ & $-2.18 \pm  0.50$ & $  68 \pm   02$ & $0.0001$ \\
V0381 Tel             & $ 1.657 \pm  0.042$ &  $3.8304$ & $-0.181 \pm  0.003$ & $ 1.649 \pm  0.019$ & $-1.18 \pm  0.11$ & $  40 \pm   15$ & $0.0003$ \\
AQ    Leo             & $ 1.783 \pm  0.037$ &  $3.8192$ & $-0.117 \pm  0.010$ & $ 1.743 \pm  0.030$ & $-1.88 \pm  0.35$ & $  56 \pm   28$ & $0.0002$ \\
V5644 Sgr             & $ 1.826 \pm  0.054$ &  $3.8283$ & $-0.191 \pm  0.004$ & $ 1.623 \pm  0.025$ & $-1.04 \pm  0.12$ & $  48 \pm   00$ & $0.0060$ \\
NN    Boo             & $ 1.655 \pm  0.031$ &  $3.8268$ & $-0.176 \pm  0.001$ & $ 1.646 \pm  0.021$ & $-1.18 \pm  0.11$ & $  30 \pm   14$ & $0.0069$ \\
BS    Com             & $ 1.693 \pm  0.042$ &  $3.8327$ & $-0.148 \pm  0.013$ & $ 1.711 \pm  0.032$ & $-1.70 \pm  0.30$ & $  59 \pm   06$ & $0.0002$ \\
V0363 Dra             & $ 1.701 \pm  0.033$ &  $3.8220$ & $-0.127 \pm  0.011$ & $ 1.739 \pm  0.031$ & $-1.88 \pm  0.32$ & $  60 \pm   08$ & $0.0002$ \\
SW    Ret             & $ 1.487 \pm  0.043$ &  $3.8217$ & $-0.166 \pm  0.007$ & $ 1.635 \pm  0.015$ & $-1.18 \pm  0.05$ & $  27 \pm   00$ & $0.0237$ \\
AL    Vol             & $ 1.690 \pm  0.027$ &  $3.8274$ & $-0.129 \pm  0.020$ & $ 1.735 \pm  0.005$ & $-1.88 \pm  0.00$ & $  59 \pm   25$ & $0.0003$ \\
V0374 Tel             & $ 1.795 \pm  0.066$ &  $3.8243$ & $-0.172 \pm  0.003$ & $ 1.643 \pm  0.019$ & $-1.18 \pm  0.11$ & $  28 \pm   08$ & $0.0122$ \\
AZ    For             & $ 1.725 \pm  0.036$ &  $3.8182$ & $-0.125 \pm  0.008$ & $ 1.772 \pm  0.015$ & $-2.40 \pm  0.26$ & $  68 \pm   05$ & $0.0002$ \\
CZ    Phe             & $ 1.720 \pm  0.037$ &  $3.8196$ & $-0.104 \pm  0.021$ & $ 1.772 \pm  0.005$ & $-2.18 \pm  0.15$ & $  62 \pm   11$ & $0.0002$ \\
V0338 Boo             & $ 1.669 \pm  0.040$ &  $3.8207$ & $-0.174 \pm  0.005$ & $ 1.643 \pm  0.018$ & $-1.18 \pm  0.07$ & $  26 \pm   17$ & $0.0119$ \\
V0458 Her             & $ 1.880 \pm  0.045$ &  $3.8253$ & $-0.169 \pm  0.002$ & $ 1.657 \pm  0.020$ & $-1.28 \pm  0.11$ & $  21 \pm   16$ & $0.0078$ \\
XY    Crv             & $ 1.803 \pm  0.044$ &  $3.8309$ & $-0.174 \pm  0.001$ & $ 1.678 \pm  0.021$ & $-1.40 \pm  0.13$ & $  60 \pm   23$ & $0.0002$ \\
V0416 Pav             & $ 1.586 \pm  0.031$ &  $3.8204$ & $-0.164 \pm  0.006$ & $ 1.647 \pm  0.016$ & $-1.18 \pm  0.05$ & $  26 \pm   01$ & $0.0175$ \\
CU    Com             & $ 1.794 \pm  0.057$ &  $3.8207$ & $-0.121 \pm  0.011$ & $ 1.741 \pm  0.013$ & $-1.88 \pm  0.09$ & $  58 \pm   29$ & $0.0003$ \\
CR    Cap             & $ 1.636 \pm  0.067$ &  $3.8281$ & $-0.164 \pm  0.009$ & $ 1.661 \pm  0.029$ & $-1.40 \pm  0.18$ & $  02 \pm   11$ & $0.0068$ \\
AG    PsA             & $ 1.635 \pm  0.045$ &  $3.8305$ & $-0.119 \pm  0.001$ & $ 1.735 \pm  0.021$ & $-1.88 \pm  0.15$ & $  54 \pm   09$ & $0.0004$ \\
V2493 Oph             & $ 1.657 \pm  0.042$ &  $3.8340$ & $-0.168 \pm  0.008$ & $ 1.670 \pm  0.015$ & $-1.40 \pm  0.05$ & $  28 \pm   18$ & $0.0003$ \\
BN    UMa             & $ 1.711 \pm  0.059$ &  $3.8166$ & $-0.144 \pm  0.007$ & $ 1.697 \pm  0.027$ & $-1.54 \pm  0.17$ & $  27 \pm   13$ & $0.0013$ \\
ASAS  J211848-3430.4  & $ 1.729 \pm  0.063$ &  $3.8150$ & $-0.145 \pm  0.008$ & $ 1.657 \pm  0.015$ & $-1.40 \pm  0.06$ & $  24 \pm   02$ & $0.0275$ \\
V0633 Cen             & $ 1.707 \pm  0.062$ &  $3.8400$ & $-0.182 \pm  0.017$ & $ 1.706 \pm  0.008$ & $-1.54 \pm  0.00$ & $  72 \pm   09$ & $0.0002$ \\
QW    Aqr             & $ 1.663 \pm  0.075$ &  $3.8328$ & $-0.174 \pm  0.005$ & $ 1.677 \pm  0.017$ & $-1.40 \pm  0.07$ & $  60 \pm   32$ & $0.0002$ \\
ASAS  J040054-4923.8  & $ 1.718 \pm  0.050$ &  $3.8196$ & $-0.134 \pm  0.001$ & $ 1.740 \pm  0.021$ & $-1.88 \pm  0.24$ & $  64 \pm   08$ & $0.0003$ \\
ASAS  J141539+0010.1  & $ 1.674 \pm  0.072$ &  $3.8318$ & $-0.155 \pm  0.016$ & $ 1.693 \pm  0.035$ & $-1.54 \pm  0.30$ & $  54 \pm   20$ & $0.0002$ \\
CF    Del             & $ 1.832 \pm  0.105$ &  $3.8250$ & $-0.172 \pm  0.002$ & $ 1.646 \pm  0.020$ & $-1.18 \pm  0.11$ & $  29 \pm   04$ & $0.0103$ \\
\hline
\end{tabular}}
\end{minipage}
\begin{flushleft}
\vspace{-5pt}
{\bf Notes:}
-- Except for the Gaia luminosities ($L_G$) all errors show the variations caused by a change of $\pm 0.005$ in $\log T_{\rm eff}$ as given in this table.
-- Mass and luminosity are in solar units, metal  abundance [M/H] is given for $Z_{\odot} = 0.0152$, age is in Myr, and counted from the ZAHB.
-- The last column shows the distance of the closest points between the pulsation and evolutionary models in the $\log L - \log M$ plane (see Eq.~\ref{eq_ml_match}).
\end{flushleft}
\end{table*}

We see that the two independent sets of luminosities (Gaia and 
RRd-based) correlate rather remarkably well if we deselect 
the six outliers marked by red squares. Although parallax errors 
undoubtedly are good candidates to contribute to their outlier 
status, it is worthwhile to examine other factors, such as 
blending (see also Sect.~\ref{sect:Vmag}). It turns out that 
all six objects suffer from various degrees of crowding -- at 
least relative to the resolutions of the instruments contributing 
to the datasets used.\footnote{The resolution limits for ASAS, 
WISE and Gaia are $15"$, $\sim 6"$ and $\sim 2"$, respectively 
-- see Sect.~\ref{sect:Vmag} and \cite{lindegren2021,ren2021}.} 
Although for CF~Del and V0374~Tel we use the transformed Gaia BP 
magnitudes, their outlier status could not be eliminated. For 
the remaining four stars (V0458~Her, V5644~Sgr, SW~Ret and 
XY~Crv) we find that the ASAS photometry yields values quite 
close to the transformed Gaia BP magnitudes, so we use the ASAS 
data. After examining the WISE image 
stamps\footnote{\url{https://irsa.ipac.caltech.edu/applications/wise/}} 
for the six outliers, we find that the WISE fluxes for SW~Ret, 
XY~Crv and V0374~Tel might have been affected by close companions, 
if the aperture was shifted toward the photocenters. 

In seeking for blend-related cause of the outlier status, it 
is worthwhile to recall that any systematic change in <V> will 
affect both the Gaia and the RRd luminosities in a similar way. 
This is because we use $V$ and $K$ (via the unWISE magnitudes) 
to estimate $T_{\rm eff}$. If we increase $V$ (e.g., using Gaia 
photometry to cure blending), then we also increase $V-K$ 
($K$ remains the same). As a result, $T_{\rm eff}$ becomes 
lower, yielding lower luminosity from the RRd analysis (see 
Fig.~3 of KW99). The higher <V>, of course, results in a lower 
luminosity for the same Gaia parallax. Therefore, the status 
(outlier or not) does not change much by introducing changes 
in <V>. The situation is different if only the infrared fluxes 
are influenced by blending. If corrected, this will lead to 
fainter $K$ magnitude and therefore, higher $T_{\rm eff}$, that 
is, higher RRd luminosity, i.e., weakening the outlier status. 
Unfortunately, this may work only for the three WISE blend 
candidates listed above. We conclude that it is unclear at 
this point what is the exact underlying cause of the outlier 
status of these six stars. 

The derived stellar parameters of the $30$ RRd stars are displayed 
in Table~\ref{tab_rrd_par}. The metric used to find the parameters 
yielding the closest match between the pulsation and evolutionary 
models (Eq.~\ref{eq_ml_match}) is also given. There are $11$ stars 
with weak matching metric, meaning basically lack of solution 
(i.e., lack of crossing the LNA and HBEV curves -- see 
Fig.~\ref{fig_ML_match}). This issue can be remedied by an increase 
of the $T_{\rm eff}$ ZP.\footnote{The temperature increase shifts 
the LNA lines to higher values in the $M$--$L$ plane with an 
opposite effect on the HBEV curves, leading to a better chance 
for intersecting.} Indeed, with a ZP shift of 
$\Delta \log T_{\rm eff}=+0.005$ leaves only $3$ stars with 
relative large $DML$ values of $0.01-0.02$. It is also noted 
that only some of the outliers discussed above have this type 
of low fidelity solution.

Using stellar evolution models with $\alpha$ enhancement may 
also have some effect on the solution but it is unlikely that 
it acts toward the improvement of the solution. With the same 
total heavy element abundance, $\alpha$ enhanced models have 
lower $\log L$ by $\sim 0.02$ \citep{dorman1992,vandenberg2000}. 
For the pulsation models the $\alpha$ enhancement also acts 
in lowering both the mass and the luminosity at the same level 
\citep{kovacs1992,kovacs1999}, so the net effect might be rather 
small.  
%
%
\begin{figure}[h]
\centering
\includegraphics[width=0.40\textwidth]{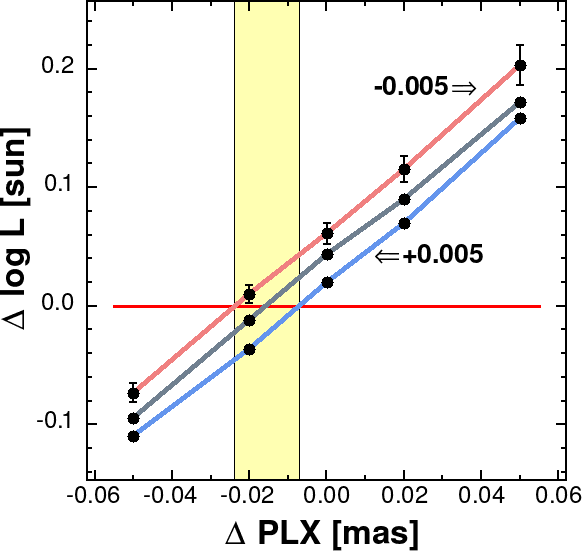}
\caption{Dependence of the average luminosity difference 
         $\log({\rm Gaia})-\log({\rm RRd})$ on the 
	 parallax shift (published minus shifted) for the 
	 RRd star sample. The separate lines correspond 
	 to $\log T_{\rm eff}$ ZP shifts (relative to our 
	 adopted ZP as given in Sect.~\ref{sect:teff_zp}) 
	 of $-0.005$, $0.0$ and $+0.005$. Error bars show 
	 the statistical errors of the mean differences. 
	 To avoid jamming, the error bars are shown only 
	 for one $T_{\rm eff}$ ZP shift (others are very 
	 similar). Requesting exact equality of the luminosity 
	 averages, the shaded area indicates the range of 
	 the allowed EDR3 parallax correction.}
\label{fig_ZP}
\end{figure}

Considering the still existing ambiguities in the temperature 
scale (\citealp{boyajian2013} -- see, however, \citealp{casagrande2014}), 
it is important to examine how the systematic shift in the 
Gaia parallax (as it seems necessary to employ at this point) 
changes as we change ZP of $T_{\rm eff}$. The ZP of the 
IRFM/GB09-adjusted $\log T_{\rm eff}$ scale we use is $3.9113$ 
(see Eq.~\ref{eq_teff} and subsequent text). We test three ZP 
shifts, $-0.005$, $0.0$ and $+0.005$ with respect to this ZP. 
For each ZP shift we scan the average difference between the Gaia 
and RRd luminosities as a function of the parallax shift with 
respect of the published EDR3 values. We see from Fig.~\ref{fig_ZP} 
that if we accept the ZP dictated by the IRFM work of 
\cite{gonzalez2009}, then adding $0.02$~mas to the published 
parallaxes can be considered as appropriate to bring the Gaia 
luminosities into agreement with the RRd luminosities. On the 
other hand, with an increase of $\Delta\log T_{\rm eff}=0.005$ 
we may no need to change anything with the EDR3 parallaxes. 
We recall (as just mentioned) that with a higher temperature 
scale the match between the LNA and HBEV models becomes also 
much better. 

Studies on various distance indicators show that the systematic 
bias in the published Gaia parallaxes decreased at each step 
of the new releases. From the period-luminosity relation of a
large sample of W~Uma binaries, \cite{ren2021} derive that an 
overall shift of $0.029$~mas of the EDR3 parallaxes (that is, 
{\em adding} $0.029$~mas to the published parallaxes) brings 
them into agreement with those obtained from the standard 
binary star analyses. The shift differs only by $0.004$~mas 
from the one suggested by the Gaia team \citep{lindegren2021}. 
Earlier, by using the large sample of benchmark eclipsing 
binaries of \cite{stassun2016}, the same authors 
\citep{stassun2021} landed at a similar conclusion, suggesting 
a somewhat smaller overall shift of $0.025$~mas with substantial 
variation over the ecliptic latitude. The parallax shift derived 
in this paper favors the above values. However, it is important 
to emphasize the role of the temperature zero point (as discussed 
above). Both in the RRd and in the eclipsing binary analyses, a 
crucial point is the choice of this parameter. A sufficient 
upward modification may lead to complete agreement, without any 
parallax shift.

%
%
\section{PLZ relation from RRd stars}
Because of the increasing importance of the period-luminosity-metallicity 
(PLZ) relations in the past twenty or so years in the 
field of RR~Lyrae stars, here we make a brief comparison 
between the PLZ resulting from the RRd stars derived in 
this paper for the 2MASS K band and some of the PLZs derived 
by other means. We note that in similar comparative work, 
\cite{dekany2009} employed $20$ RRd stars ($3$ from the galactic 
field and $17$ from the Large Magellanic Cloud) to compare the 
Wesenheit indices $W(V,B-V)$ of these stars with those obtained 
from the Baade-Wesselink (B-W) analysis of 22 galactic FU 
stars \citep{kovacs2003}. The conclusion of the work of 
\cite{dekany2009} was that the $P_0$--$W(V,B-V)$ relations obtained 
from the RRd analysis (akin to the one presented in this paper) 
and from the B-W analysis were statistically consistent. However, 
the RRd relation turned out to be much tighter and steeper than 
the one resulting from the B-W analysis.   

The derivation of the RRd PLZ is very straightforward. 
The RRd analysis yields $P_0$, $T_{\rm eff}$, $L$ and $Z$. 
Then, knowing these quantities, we can apply bolometric 
corrections to the luminosity values and get the visual 
absolute magnitude $M_V$. After inverting Eq.~\ref{eq_teff} 
for $M_V-M_K$, we get the absolute 2MASS $K$ magnitude simply 
by subtracting $0.03$ from $M_K$ (see Sect.~\ref{sect:Kmag}).    

\begin{figure}[h]
\centering
\includegraphics[width=0.40\textwidth]{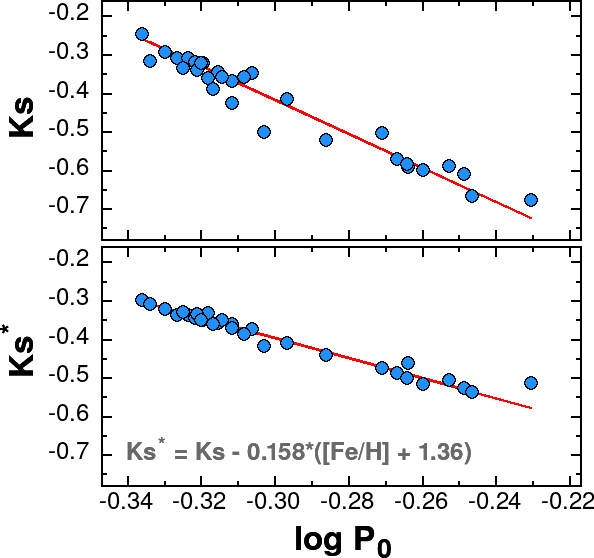}
\caption{{\em Upper panel:} Absolute Ks magnitudes derived 
         from pulsation and evolution models for the $30$ 
	 galactic RRd stars of Table~\ref{tab_obs_ref} as a 
	 function of the fundamental period. 
	 {\em Lower panel:} As above, but for the modified 
	 Ks values for the metallicity effect. 
	 Continuous lines show the respective regressions (see 
	 Eq.~\ref{eq_plz_fit} for the fit shown in the lower 
	 panel).}
\label{fig_PLZ_RRd}
\end{figure}

The resulting $\log P_0$--$K_s$ plot (where $K_s$ denotes the 
2MASS K magnitude) is shown in the upper panel of 
Fig.~\ref{fig_PLZ_RRd}. As expected, the correlation is 
significant but the scatter is somewhat excessive. It has 
been long claimed and shown in several papers 
\citep[e.g.,][]{bono2003} that there is also a significant 
metallicity dependence for $K_s$. Indeed, as shown in the 
lower panel of Fig.~\ref{fig_PLZ_RRd}, by regressing both the 
period and the metallicity, we get a considerably tighter 
correlation. With three relatively `mild' outliers (XX~Crv, 
AZ~For and AG~PsA) a robust fit yields the following formula         
%
%
\begin{eqnarray}
\label{eq_plz_fit}
K_s&=&-(0.396\pm 0.003) - (2.606\pm 0.134)(\log P_0 + 0.30) \nonumber \\
   &+&\phantom{-}(0.158\pm 0.011)([Fe/H]+1.36) \hspace{2mm}.    
\end{eqnarray}
The standard deviation of the above regression is $0.0086$~mag. 
It is interesting to examine the $T_{\rm eff}$ ZP issue 
discussed in Sect.~\ref{sect:LL} in terms of the tightness of 
the RRd PLZ relation. It seems that the tightness of this 
relation favors fairly significantly to the GB09 scale. 
By changing the ZP of $\log T_{\rm eff}$ for the three-parameter 
fits, we get residual standard deviations of $0.025$ and $0.043$ 
magnitudes for $\Delta\log T_{\rm eff}=-0.005$ and $+0.005$, 
respectively. This result seems to contradict to the higher 
temperature scale suggested by the better HBEV/LNA match accuracy 
of the models and further supports the need for the correction 
of the Gaia parallaxes.\footnote{In a current paper on single-mode 
galactic RR~Lyrae stars \cite{marconi2021} conclude that the 
parallax correction might be statistically insignificant. However, 
it seems that their data are too noisy to detect the small 
parallax shift claimed by other studies.} Nevertheless, it is 
unclear how tight the true PLZ should be. There is a hidden mass 
and temperature dependence in the PLZ relation, and it is not 
entirely known how much evolutionary and color effects make these 
parameters correlated with each other, and thereby tighten the 
observed PLZ relation.   
      
%
%

\begin{figure}[h]
\centering
\includegraphics[width=0.40\textwidth]{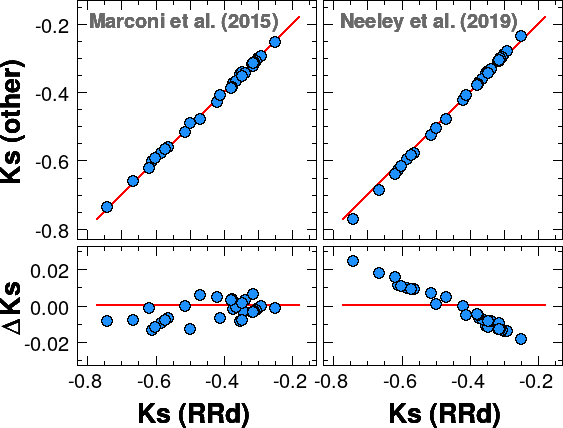}
\caption{Comparison of the PLZ relations as given in 
         Table~\ref{tab_PLZ_compare}. All formulae are compared  
	 on the set of ($\log P_0$,[Fe/H]) values given in 
	 Table~\ref{tab_rrd_par}. The lower panels show the 
	 differences 
	 $\Delta {\rm Ks} = {\rm Ks}({\rm RRd})-{\rm Ks}({\rm other})$ 
	 between the pair of formulae.}
\label{fig_PLZ_compare}
\end{figure}

In a comparison with other, recent variants of the RR~Lyrae 
PLZ relation, we choose the relation of \cite {marconi2015}, 
based on pulsation and stellar evolution considerations, and 
that of \cite{neeley2019} from the purely empirical fit to 
the bright, single-mode RR~Lyrae stars with high-fidelity 
Gaia (DR2) parallaxes. The PLZ relations are compared in 
Fig.~\ref{fig_PLZ_compare}. The agreement is very good 
between all these independent estimates with tighter but 
systematically more deviating trend for the relation based 
on the Gaia DR2 parallaxes. In a comparison of the regression 
coefficients (see Table~\ref{tab_PLZ_compare}), we see that 
although our formula seems to be closer to that of 
\cite{neeley2019} the systematic trend is stronger than 
for the formula of \cite {marconi2015}. This underlines the 
significance of the metallicity term, that is stronger in 
both of these formulae than in ours. This is partially 
compensated by the weaker period dependence in the formula 
of \cite{marconi2015} but works in the opposite way with 
the strong period dependence derived by \cite{neeley2019}. 
In spite of these details, likely due to the errors of 
the fits, we think that the overall agreement with 
standard deviations of $0.005$--$0.01$~mag is very satisfying 
between the different approximations.  

%
%
\begin{table}[h!]
\centering
\begin{minipage}{200mm}
\caption{Comparison of the PLZ regression coefficient}
\label{tab_PLZ_compare}
\scalebox{1.0}{
\begin{tabular}{cccl}
\hline
 a   & b & c & Source \\
\hline\hline
$-0.390$ & $-2.250$ & $+0.180$ & \cite{marconi2015} \\
$-0.390$ & $-2.730$ & $+0.180$ & \cite{neeley2019}  \\
$-0.396$ & $-2.606$ & $+0.158$ & this paper \\
\hline
\end{tabular}}
\end{minipage}
\begin{flushleft}
\vspace{-5pt}
{\bf Note:} 
The coefficients refer to the following type of formula: 
${\rm Ks}=a+b(\log P_0+0.30)+c({\rm [Fe/H]}+1.36)$
\end{flushleft}
\end{table}
%

%
%
%

%
%
\section{Conclusions}
The steadily improving accuracy of the Gaia parallaxes allows 
us to visit increasing number of objects and derive various physical 
parameters with impressively high precision.  
RR~Lyrae stars are among those 
objects that come into the forefront of investigation, due to their 
significance in galactic structure and stellar evolution. In this 
paper we used relatively bright galactic double-mode RR~Lyrae (RRd) 
stars and confronted their (largely) theoretical luminosities with 
those obtained (almost) directly from their parallaxes currently 
made available by the third (early) data release (EDR3) of the Gaia 
mission. With almost all individual parallax errors 
less than $10$\%, and sample size of $30$, it is possible to aim 
for few percent statistical precision in defining the zero point 
of the RR~Lyrae luminosity scale and investigate the source of 
the remaining systematic differences.  

We implemented a very similar method to those used in our earlier 
investigations. The method is based on the simple observation that 
the two periods yield a closed solution for the stellar parameters, 
assuming that we have reliable estimates on the temperature and 
metallicity. Unfortunately, for the stars under scrutiny, the 
latter quantity is unavailable, therefore we had to resort to 
the combination of pulsation and evolutionary models 
\citep[BaSTI,][]{hidalgo2018}, together with observationally 
determined temperature much in the same way as in \cite{dekany2008}. 
In return, this method yields a full solution, including an 
estimate on the metallicity. Our main results are as follows. 

\begin{itemize}
\item
Except for $6$ outlying stars (likely affected by blending and 
parallax errors), the luminosities of the remaining $24$ stars 
correlate well with those derived from the Gaia EDR3 parallaxes. 
\item
By fixing the $T_{\rm eff}$ zero point to the one given by 
\cite{gonzalez2009}, we get no systematic differences between 
the Gaia and RRd luminosities, assuming that the Gaia parallaxes 
are shifted by $0.02$~mas upward. This result is consonant with 
other studies suggesting similar (albeit some $0.005$--$0.01$~mas 
larger) corrections. 
\item
We derived a period-luminosity-metallicity (PLZ) relation for 
the 2MASS K color, and found that it is in good agreement 
with two other, independent PLZ relations. The first relation  
\citep{neeley2019} is based on direct Gaia DR2 distances, 
whereas the second \citep{marconi2015} comes from pulsation-stellar 
evolution considerations. Our PLZ relation is fundamentally 
theoretical, aided by the observationally calibrated zero 
point of $T_{\rm eff}$. There are no significant differences 
in the zero points and the predicted values have a scatter 
(standard deviation) of only $\sim 0.01$~mag.    
\item
The tightness criterion of the RRd PLZ relation significantly 
favors the $T_{\rm eff}$ zero point accepted for this study. 
This lends further support to the upward correction of the EDR3 
parallaxes as mentioned above.  
\end{itemize}

The work presented in this paper would have been much simpler, 
and independent of stellar evolutionary models, if we had reliable 
metal abundances available for our targets. By knowing 
the metallicity and the temperature, one can estimate the 
luminosity using only pulsation theory and directly compare 
these values with the Gaia luminosities, without the intermediary 
of the evolutionary models. The future availability of metallicity 
for bright double-mode variables, and, in particular, for 
RR~Lyrae stars, is an important step for a nearly direct test of  
the Gaia parallaxes and evolutionary models.

%
%
\begin{acknowledgements}
%
It is a pleasure to thank to Andy Monson for the valuable 
correspondence on the calibration of the photometric data used 
in this paper. 
%
This work is based heavily on the availability of well-calibrated 
photometry of the All Sky Automated Survey (ASAS). We thank to 
Grzegorz Pojmanski and his co-workers for running this great project 
for so many years.  
%
This research has made use of the VizieR catalogue access tool, CDS, 
Strasbourg, France (DOI: 10.26093/cds/vizier). 
%
This work has made use of data from the European Space Agency (ESA) 
mission {\it Gaia} (\url{https://www.cosmos.esa.int/gaia}), 
processed by the {\it Gaia} Data Processing and Analysis Consortium 
(DPAC, \url{https://www.cosmos.esa.int/web/gaia/dpac/consortium}). 
Funding for the DPAC has been provided by national institutions, 
in particular the institutions participating in the {\it Gaia} 
Multilateral Agreement.
%
This publication makes use of data products from the Wide-field 
Infrared Survey Explorer, which is a joint project of the University 
of California, Los Angeles, and the Jet Propulsion 
Laboratory/California Institute of Technology, funded by the 
National Aeronautics and Space Administration.
%
This research has made use of the NASA/IPAC Infrared Science Archive, 
which is funded by the National Aeronautics and Space Administration 
and operated by the California Institute of Technology.
%
This research has made use of the International Variable Star Index 
(VSX) database, operated at AAVSO, Cambridge, Massachusetts, USA.
%
Supports from the National Research, Development and Innovation 
Office (grants K~129249 and NN~129075) are acknowledged. 
\end{acknowledgements}

%
%

%

%
%
\begin{appendix}
\section{Observational datasets used in this work}
\label{app_A}
We summarize the observational data used in this study in 
Table~\ref{tab_obs_dat}. The discoveries and the periods were 
reported by the papers listed in Table~\ref{tab_obs_ref}. 
The flux-averaged magnitudes in the Johnson V filter are 
based on the 3$^{rd}$-order Fourier fit (see Sect.~\ref{sect:Vmag}) 
to the ASAS V observations by starting from the published 
periods and refining them due to the long time span of the 
ASAS data. Four entries (marked by double asterisks) do 
not have ASAS light curves, therefore their <V> values were 
approximated with the aid of the Gaia photometry, as described 
in Sect.~\ref{sect:Vmag}. The same procedure was employed 
on three more stars (V0381~Tel, V0374~Tel, CF~Del), where 
blending and sparse light curve sampling corrupted the ASAS 
magnitudes.  

For completeness, the flux-averaged Gaia EDR3 magnitudes are 
listed for all three bands, even though we use only the BP 
magnitudes. For the ASAS <V> values we take a flat error of 
$0.005$~mag for all entries, whereas for those derived from 
the Gaia average fluxes, we give the errors as given by the 
EDR3 catalog. 

The W1, W2 fluxes come from the recent unWISE catalog by 
\cite{schlafly2019} and result from the band-integrated fluxes 
throughout the full mission of the satellite. The reddening 
values are from the map of \cite{schlafly2011} accessible 
from the NASA/IPAC Infrared Science Archive. 
Parallaxes are from the Gaia EDR3 catalog (accessed from the 
VizieR database).

%
%
%
\begin{table*}[h!]
\centering
\begin{minipage}{200mm}
\caption{Observed properties of 30 Galactic RRd stars}
\label{tab_obs_dat}
\scalebox{0.90}{
\begin{tabular}{lcc rrrrrrcc}
\hline
Target & $P_0$ & $P_1/P_0$ & <V> & G & BP & RP & W1 & W2 & E(B-V) & PLX\\
       &  (d)  &  (d)  &(mag)&(mag)&(mag)&(mag)&(flux)&(flux)&(mag)&(mas)\\
\hline
V0500\_Hya      & 0.563906 & 0.746204 & 10.772 & 10.729 & 10.945 & 10.382 & 138300 & 136290 & 0.020 & 0.8453\\
                & 0.000000 & 0.000000 &  0.005 &  0.008 &  0.025 &  0.017 &     58 &    121 & 0.001 & 0.0264\\
V0372\_Ser      & 0.471349 & 0.744098 & 11.320 & 11.268 & 11.492 & 10.872 &  95313 &  93210 & 0.073 & 0.8083\\
                & 0.000000 & 0.000000 &  0.005 &  0.011 &  0.041 &  0.025 &     49 &    103 & 0.008 & 0.0204\\
Z\_Gru          & 0.487995 & 0.744243 & 12.340 & 12.281 & 12.450 & 11.958 &  30145 &  29373 & 0.022 & 0.4794\\
                & 0.000000 & 0.000000 &  0.005 &  0.011 &  0.035 &  0.022 &     28 &     63 & 0.001 & 0.0170\\
XX\_Crv         & 0.544515 & 0.746429 & 12.375 & 12.282 & 12.468 & 11.921 &  32925 &  32660 & 0.058 & 0.4220\\
                & 0.000000 & 0.000000 &  0.005 &  0.007 &  0.025 &  0.016 &     30 &     67 & 0.004 & 0.0156\\
V0381\_Tel$^{**}$ & 0.467819 & 0.743418 & 12.782 & 12.711 & 12.887 & 12.378 &  20806 &  20358 & 0.041 & 0.3724\\
                & 0.000000 & 0.000000 &  0.032 &  0.009 &  0.032 &  0.020 &     25 &     56 & 0.001 & 0.0178\\
AQ\_Leo         & 0.549751 & 0.746063 & 12.530 & 12.470 & 12.662 & 12.119 &  27604 &  26799 & 0.022 & 0.3571\\
                & 0.000000 & 0.000000 &  0.005 &  0.009 &  0.030 &  0.019 &     28 &     63 & 0.001 & 0.0155\\
V5644\_Sgr      & 0.461258 &  0.74248 & 12.546 & 12.465 & 12.701 & 12.064 &  33528 &  32767 & 0.130 & 0.3884\\
                & 0.000000 & 0.000000 &  0.005 &  0.012 &  0.042 &  0.026 &     31 &     69 & 0.005 & 0.0249\\
NN\_Boo$^{**}$  & 0.474600 & 0.743742 & 12.683 & 12.619 & 12.790 & 12.304 &  21747 &  21172 & 0.013 & 0.3760\\
                & 0.000000 & 0.000000 &  0.025 &  0.008 &  0.025 &  0.016 &     24 &     53 & 0.001 & 0.0130\\
BS\_Com         & 0.487902 & 0.744121 & 12.734 & 12.672 & 12.842 & 12.370 &  19897 &  19455 & 0.012 & 0.3513\\
                & 0.000000 & 0.000000 &  0.005 &  0.007 &  0.023 &  0.014 &     23 &     53 & 0.000 & 0.0175\\
V0363\_Dra$^{**}$ & 0.540800 & 0.745562 & 12.768 & 12.672 & 12.874 & 12.342 &  21960 &  21378 & 0.027 & 0.3533\\
                & 0.000000 & 0.000000 &  0.027 &  0.008 &  0.027 &  0.017 &     23 &     49 & 0.000 & 0.0133\\
SW\_Ret         & 0.476624 & 0.744425 & 12.805 & 12.727 & 12.948 & 12.375 &  22989 &  22492 & 0.060 & 0.4681\\
                & 0.000000 & 0.000000 &  0.005 &  0.006 &  0.020 &  0.012 &     24 &     52 & 0.006 & 0.0234\\
AL\_Vol         & 0.517218 & 0.744781 & 12.806 & 12.691 & 12.914 & 12.306 &  25243 &  24770 & 0.107 & 0.3951\\
                & 0.000000 & 0.000000 &  0.005 &  0.007 &  0.025 &  0.015 &     25 &     53 & 0.003 & 0.0126\\
V0374\_Tel$^{**}$ & 0.479065 & 0.743866 & 13.193 & 13.102 & 13.294 & 12.737 &  15586 &  15079 & 0.053 & 0.2628\\
                & 0.000000 & 0.000000 &  0.045 &  0.012 &  0.044 &  0.026 &     22 &     50 & 0.001 & 0.0203\\
AZ\_For         & 0.588230 & 0.745627 & 12.894 & 12.814 & 13.020 & 12.489 &  19462 &  18999 & 0.012 & 0.3178\\
                & 0.000000 & 0.000000 &  0.005 &  0.006 &  0.019 &  0.012 &     23 &     52 & 0.000 & 0.0137\\
CZ\_Phe         & 0.566809 & 0.745392 & 12.898 & 12.812 & 13.010 & 12.490 &  19153 &  18580 & 0.011 & 0.3178\\
                & 0.000000 & 0.000000 &  0.005 &  0.006 &  0.021 &  0.013 &     23 &     51 & 0.001 & 0.0140\\
V0338\_Boo$^{**}$ & 0.494040 & 0.742673 & 12.959 & 12.860 & 13.063 & 12.546 &  17847 &  17309 & 0.015 & 0.3258\\
                & 0.000000 & 0.000000 &  0.031 &  0.009 &  0.031 &  0.020 &     22 &     47 & 0.001 & 0.0149\\
V0458\_Her      & 0.483740 &  0.74416 & 13.123 & 13.027 & 13.233 & 12.681 &  16715 &  16213 & 0.058 & 0.2470\\
                & 0.000000 & 0.000000 &  0.005 &  0.007 &  0.021 &  0.013 &     22 &     49 & 0.002 & 0.0136\\
XY\_Crv         & 0.484820 & 0.743851 & 13.311 & 13.247 & 13.445 & 12.872 &  14432 &  13971 & 0.082 & 0.2565\\
                & 0.000000 & 0.000000 &  0.005 &  0.008 &  0.028 &  0.018 &     21 &     50 & 0.003 & 0.0137\\
V0416\_Pav      & 0.491515 &  0.74411 & 13.347 & 13.259 & 13.485 & 12.859 &  15270 &  15066 & 0.091 & 0.3348\\
                & 0.000000 & 0.000000 &  0.005 &  0.006 &  0.022 &  0.014 &     21 &     49 & 0.004 & 0.0124\\
CU\_Com         & 0.544164 & 0.745659 & 13.363 & 13.227 & 13.406 & 12.907 &  12603 &  12232 & 0.020 & 0.2324\\
                & 0.000000 & 0.000000 &  0.005 &  0.008 &  0.028 &  0.018 &     19 &     47 & 0.001 & 0.0161\\
CR\_Cap         & 0.473022 & 0.744283 & 13.366 & 13.290 & 13.464 & 12.960 &  12249 &  12013 & 0.036 & 0.2866\\
                & 0.000000 & 0.000000 &  0.005 &  0.009 &  0.030 &  0.019 &     20 &     50 & 0.001 & 0.0230\\
AG\_PsA         & 0.497841 & 0.745248 & 13.379 & 13.342 & 13.522 & 13.027 &  11494 &  11130 & 0.018 & 0.2786\\
                & 0.000000 & 0.000000 &  0.005 &  0.009 &  0.034 &  0.022 &     19 &     46 & 0.002 & 0.0151\\
V2493\_Oph      & 0.463324 & 0.742977 & 13.482 & 13.403 & 13.661 & 12.980 &  14068 &  13720 & 0.140 & 0.3079\\
                & 0.000000 & 0.000000 &  0.005 &  0.008 &  0.028 &  0.017 &     20 &     47 & 0.009 & 0.0156\\
BN\_UMa$^{**}$  & 0.535786 & 0.745932 & 13.580 & 13.504 & 13.677 & 13.172 &  10359 &  10168 & 0.014 & 0.2292\\
                & 0.000000 & 0.000000 &  0.030 &  0.009 &  0.030 &  0.019 &     18 &     43 & 0.002 & 0.0162\\
J211848-3430.4  & 0.504860 & 0.745486 & 13.653 & 13.597 & 13.870 & 13.156 &  12492 &  12231 & 0.101 & 0.2472\\
                & 0.000000 & 0.000000 &  0.005 &  0.009 &  0.032 &  0.020 &     20 &     48 & 0.004 & 0.0190\\
V0633\_Cen      & 0.480517 & 0.743739 & 13.621 & 13.588 & 13.790 & 13.233 &   9865 &   9738 & 0.076 & 0.2449\\
                & 0.000000 & 0.000000 &  0.005 &  0.012 &  0.044 &  0.027 &     17 &     43 & 0.002 & 0.0184\\
QW\_Aqr         & 0.477237 & 0.743836 & 13.705 & 13.715 & 13.949 & 13.308 &  10235 &  10038 & 0.098 & 0.2564\\
                & 0.000000 & 0.000000 &  0.005 &  0.010 &  0.036 &  0.024 &     18 &     45 & 0.006 & 0.0235\\
J040054-4923.8  & 0.558588 & 0.745934 & 13.824 & 13.768 & 13.977 & 13.453 &   7994 &   7825 & 0.007 & 0.1992\\
                & 0.000000 & 0.000000 &  0.005 &  0.008 &  0.028 &  0.017 &     15 &     35 & 0.001 & 0.0123\\
J141539+0010.1  & 0.481932 & 0.744591 & 13.873 & 13.737 & 13.882 & 13.419 &   7552 &   7263 & 0.035 & 0.2114\\
                & 0.000000 & 0.000000 &  0.005 &  0.010 &  0.035 &  0.023 &     16 &     40 & 0.001 & 0.0187\\
CF\_Del$^{**}$  & 0.478448 &  0.74411 & 14.348 & 14.217 & 14.436 & 13.845 &   5880 &   5709 & 0.087 & 0.1471\\
                & 0.000000 & 0.000000 &  0.030 &  0.009 &  0.030 &  0.019 &     14 &     38 & 0.001 & 0.0196\\
\hline
\end{tabular}}
\end{minipage}
\begin{flushleft}
\vspace{-5pt}

{\bf Notes:} 
The lines below the entry lines show the corresponding errors. 
The <V> values derived from the calibration of the EDR3 BP 
magnitudes (see Eq.~\ref{eq_BP2V}) are marked by $^{**}$. 
\end{flushleft}
\end{table*}

\end{appendix}

\end{document}